\documentclass[a4paper,twocolumn,citeautoscript,prl]{revtex4-1}

\usepackage[T1]{fontenc}
\usepackage{times}
\usepackage{mathptmx}
\usepackage{amsmath}
\usepackage[dvips]{graphicx,color}
\usepackage[a4paper,body={18cm,25.6cm},top=2cm]{geometry}
\usepackage{titlesec}

\setcitestyle{super}

\titleformat{\section}{\bfseries}{\thesection}{0em}{}	%section style
\titleformat{\subsection}[runin]{\normalfont\bfseries}{}{0em}{}	%subsection style
\titlespacing*{\section}{0pt}{1em}{-\parskip}

\makeatletter
\let\bibsection\rtx@bibsection                               % Title for ref section instead of ruler
\def\fnum@figure{\textbf{Figure~\thefigure} $|$\ \@gobble}   % Nature style figure caption number
\makeatother

\begin{document}

\begin{titlepage}

%\vspace*{3cm}

\begin{center}
\LARGE{\textbf{Temporal tweezing of light:\\ trapping and manipulation of temporal cavity solitons}}

\vspace*{6mm}

%AUTHORS:
\large{Jae K. Jang, Miro~Erkintalo, St\'ephane Coen$^\text{*}$ \& Stuart~G.~Murdoch} \vspace*{3mm}

%AFFILIATIONS:
\textit{\small{
Department of Physics, University of Auckland, Private Bag 92019, Auckland, New Zealand.\\
Corresponding author: s.coen@auckland.ac.nz}}
\end{center}
\vspace*{1mm}

%ABSTRACT:
\begin{quote}
\noindent Optical tweezers use laser light to trap and move microscopic particles in space. Here we demonstrate a
similar control over ultrashort light pulses, but in time. Our experiment involves temporal cavity solitons that
are stored in a passive loop of optical fiber pumped by a continuous-wave ``holding'' laser beam. The cavity
solitons are trapped into specific time slots through a phase-modulation of the holding beam, and moved around in
time by manipulating the phase profile. We report both continuous and discrete manipulations of the temporal
positions of picosecond light pulses, with the ability to simultaneously and independently control several pulses
within a train. We also study the transient drifting dynamics and show complete agreement with theoretical
predictions. Our study demonstrates how the unique particle-like characteristics of cavity solitons can be
leveraged to achieve unprecedented control over light. These results could have significant ramifications for
optical information processing.
\end{quote}
%\endgroup
%\newpage
\vspace*{6mm}
\end{titlepage}

\setlength{\parindent}{15pt}
\setlength{\parskip}{0pt}

\noindent All-optical trapping and manipulation of the temporal positions of light pulses is a highly desirable
functionality, with immediate ramifications for optical information processing \cite{firth_cavity_2002,
lugiato_introduction_2003, boyd_applications_2006, hau_optical_2008}. Information represented as a sequence of pulses
could be stored and reconfigured on the fly, without the need for power-hungry optoelectronic conversion. This calls
for the ability to trap ultrashort pulses of light, and dynamically move them around in time, with respect to, and
independently of each other. Slow-light \cite{hau_light_1999, okawachi_tunable_2005, mok_dispersionless_2006,
thevenaz_slow_2008} and nonlinear cross-phase modulation effects \cite{rothenberg_intrafiber_1990,
de_sterke_optical_1992, nishizawa_ultrafast_2003, gorbach_light_2007, philbin_fiber-optical_2008,
webb_nonlinear_2014} can partly achieve this feat, yet neither of these approaches are sufficiently flexible to
enable independent dynamical control of light pulses within a sequence.

Enter temporal cavity solitons (CSs) \cite{wabnitz_suppression_1993, leo_temporal_2010, leo_dynamics_2013,
jang_ultraweak_2013, jang_observation_2014}. These are the dissipative solitons \cite{akhmediev_dissipative_2008,
grelu_dissipative_2012} of externally-driven nonlinear passive cavities. Specifically, they are pulses of light that
can persist indefinitely in a passive loop of nonlinear optical material such as fibre rings and monolithic
microresonators, without changing shape or losing power. Dispersive temporal spreading is arrested by the material
nonlinearity, and they draw the power they need from a continuous-wave (cw) ``holding'' laser beam driving the
cavity. As multiple CSs can be present simultaneously and independently, at arbitrary temporal positions, they
constitute ideal bits for all-optical buffer applications \cite{leo_temporal_2010}. Furthermore, their ``plasticity''
provide a solution to the problem of selective positioning control. Specifically, any gradient on the cavity holding
beam is expected to cause an overlapping CS to move towards --- and be trapped at --- a point where the gradient
vanishes. Dynamical control of these external gradients then shift the CSs. This concept has been investigated
theoretically in the context of two-dimensional \textit{spatial} CSs, beams of light persisting in the transverse
plane of planar cavities \cite{firth_optical_1996-1, firth_theory_2001, firth_cavity_2002, lugiato_introduction_2003,
spinelli_spatial_1998, maggipinto_cavity_2000}. Although experiments have had some success in positioning and moving
spatial CSs in space, local defects and non-uniformities across the plane of the cavity often somewhat restrict this
control \cite{taranenko_spatial_1997, barland_cavity_2002, pedaci_positioning_2006, gutlich_dynamic_2007,
pedaci_all-optical_2008, cleff_gradient_2008, caboche_microresonator_2009}. With \textit{temporal} CSs, material
imperfections are not an issue: Every CS circulating the cavity sees the same averaged environment
\cite{firth_temporal_2010}. Integration with existing fibre-optic communication technologies is also more natural.

Here we report on the experimental realization of trapping and selective manipulation of temporal cavity solitons.
In our experiment, the CSs exist as picosecond pulses of light, recirculating in a loop of optical fibre, and we
expose them to temporal control gradients in the form of a gigahertz phase modulation imposed on the cavity holding
beam. We show theoretically and experimentally that CSs are attracted and trapped to phase maxima, which effectively
suppresses all environmental fluctuations and soliton interactions. By dynamically changing the phase pattern, we
then controllably move the CSs in time, in essence selectively speeding them up or slowing them down. Continuous and
discrete manipulations are demonstrated, both with temporal shifts much larger than the CS duration. Additionally,
we investigate the transient CS attraction dynamics, and show complete agreement with theoretical predictions.

Our results demonstrate that individual ultrashort light pulses can be shifted temporally, forward or backward,
simply using cw laser light. This compares with conventional optical tweezers that trap and move microscopic
particles in space \cite{ashkin_observation_1986, ashkin_history_2000, grier_revolution_2003}, only here we
manipulate light itself, and in time. We therefore refer to our technique as the temporal tweezing of light.

\section*{Results}

\subsection{Concept of temporal tweezing.}

\begin{figure*}[t]
  \includegraphics[width = 0.85\textwidth]{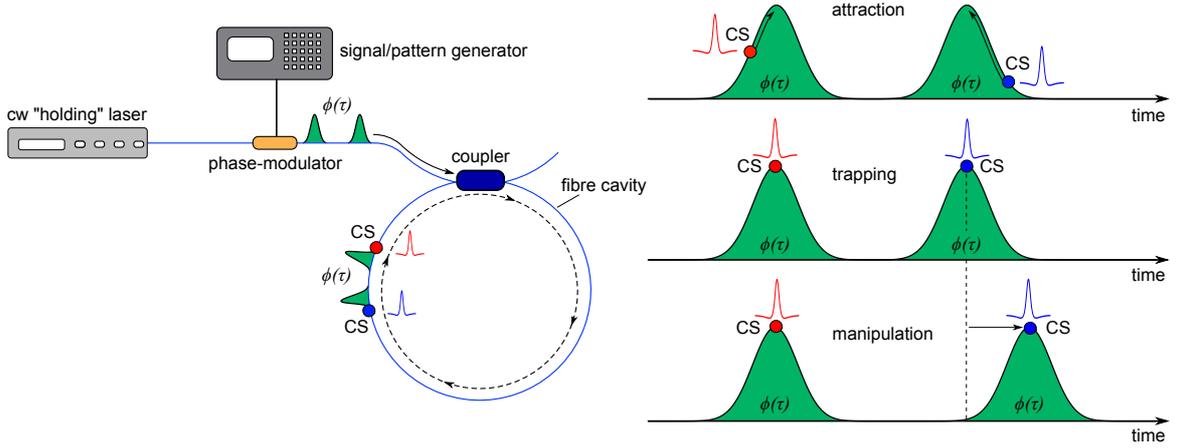}
  \caption{\small \textbf{Principle of the temporal tweezing of light.} A cw laser with an imprinted phase pattern
    $\phi(\tau)$ (green) drives a passive fibre cavity. When optical temporal CSs (red and blue) circling the cavity
    overlap with a phase gradient, their velocity changes and they drift towards the nearest phase maxima (top). The
    maxima act as temporal trapping sites for the CSs (middle). Once trapped, the CSs follow changes in the phase
    pattern: modifying the relative delay of two phase maxima enables arbitrary manipulation of the temporal delay of
    two CSs (bottom).}
  \label{schematic}
\end{figure*}
Figure~\ref{schematic} illustrates the principle underlying temporal tweezing. We consider a passive resonator
constructed of single-mode optical fibre, driven with monochromatic laser light. A modulator imprints a time-varying
electric signal $\phi(\tau)$ onto the phase of the cw holding laser driving the cavity. Provided that $\phi(\tau)$
repeats periodically, with a period that is an integer fraction of the cavity roundtrip time, the holding  beam
builds up over several roundtrips an intracavity cw field with an identical travelling phase pattern (Supplementary
Section~SI). Being effectively synchronized to the phase modulation, the intracavity phase pattern is coherently
reinforced each roundtrip, ensuring steady-state operation.

Temporal CSs circulating around the cavity are superimposed onto the cw intracavity field. They normally have the
same frequency and group-velocity as that field \cite{leo_temporal_2010}. However, the presence of a phase gradient
across a light pulse is equivalent to an instantaneous frequency shift, $\delta \omega =
-\mathrm{d}\phi/\mathrm{d}\tau=-\phi'$. Specifically, a decrease (respectively, increase) of phase over time leads to
a blue-shift (red-shift). In a material exhibiting anomalous dispersion (a necessary condition for temporal CSs to
exist  \cite{leo_temporal_2010}), this shift translates into an increase (respectively, decrease) of the
group-velocity. Accordingly, the temporal CSs catch up with the imprinted phase pattern (or vice versa), leading to
an effective time-domain drift of the CSs towards the nearest phase maxima. Over one roundtrip, the CSs accrue an
extra group delay given by (see also Supplementary Section~SI):
\begin{equation}
  \tau_\mathrm{drift} = \beta_2 L \delta\omega = |\beta_2| L \phi',
  \label{drift}
\end{equation}
where $L$ is the length of the fibre loop and $\beta_2$ is the group-velocity dispersion coefficient, and we have
assumed anomalous dispersion ($\beta_2<0$). It is clear that a CS overlapping with the leading (trailing) edge of the
phase profile will experience a positive (negative) delay; both scenarios bring the soliton closer to the nearest
maximum of the phase profile. At that point, the phase does not vary, hence the drift ceases, and the CSs propagate
at the same velocity as the phase peaks: the two do not move further with respect to each other. The phase maxima
therefore correspond to stable equilibria, acting as trapping sites for temporal CSs. Significantly, if the phase
pattern is adjusted, the CSs will follow. This means that, when multiple CSs are trapped to distinct phase peaks,
manipulating the phase of the holding beam enables ultrashort optical CSs to be freely moved in time with respect to
each other, i.e. to realize a temporal tweezer of light.

\begin{figure*}[t]
  \includegraphics[width = 0.85\textwidth]{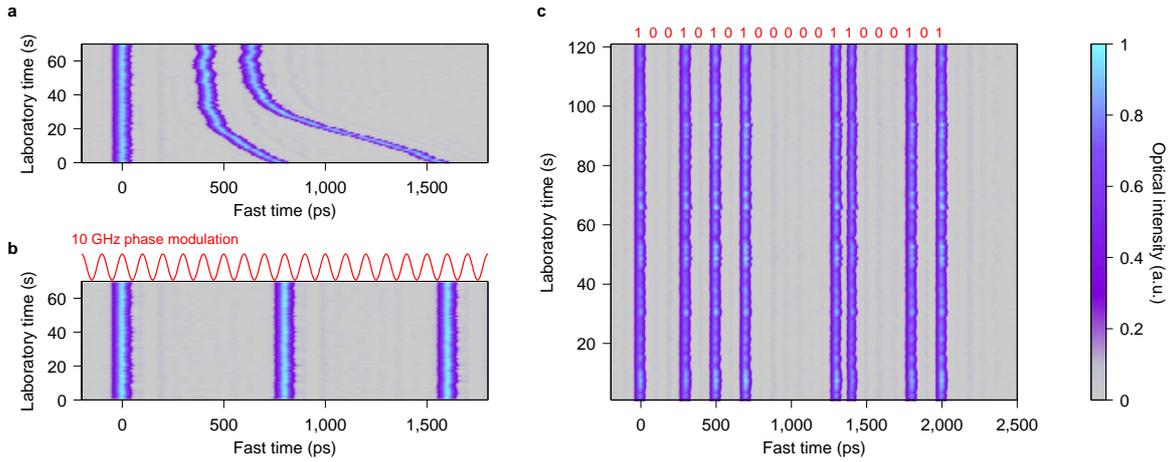}
  \caption{\small \textbf{Experimental demonstration of cavity soliton trapping.} \textbf{a,} In the absence of any
    phase modulation on the cavity holding beam, three CSs drift in time due to acoustic-mediated
    interactions. \textbf{b,} The interactions are overcome, and the CSs trapped, when applying a sinusoidal phase
    modulation at 10~GHz (shown in red on top). \textbf{c,} Phase-modulation enables all-optical data storage:
    A binary-encoded sequence of CSs at 10~Gbit/s is held for two minutes, without distortion.}
  \label{PM1}
\end{figure*}
\begin{figure*}[t]
  \includegraphics[width = 0.85\textwidth]{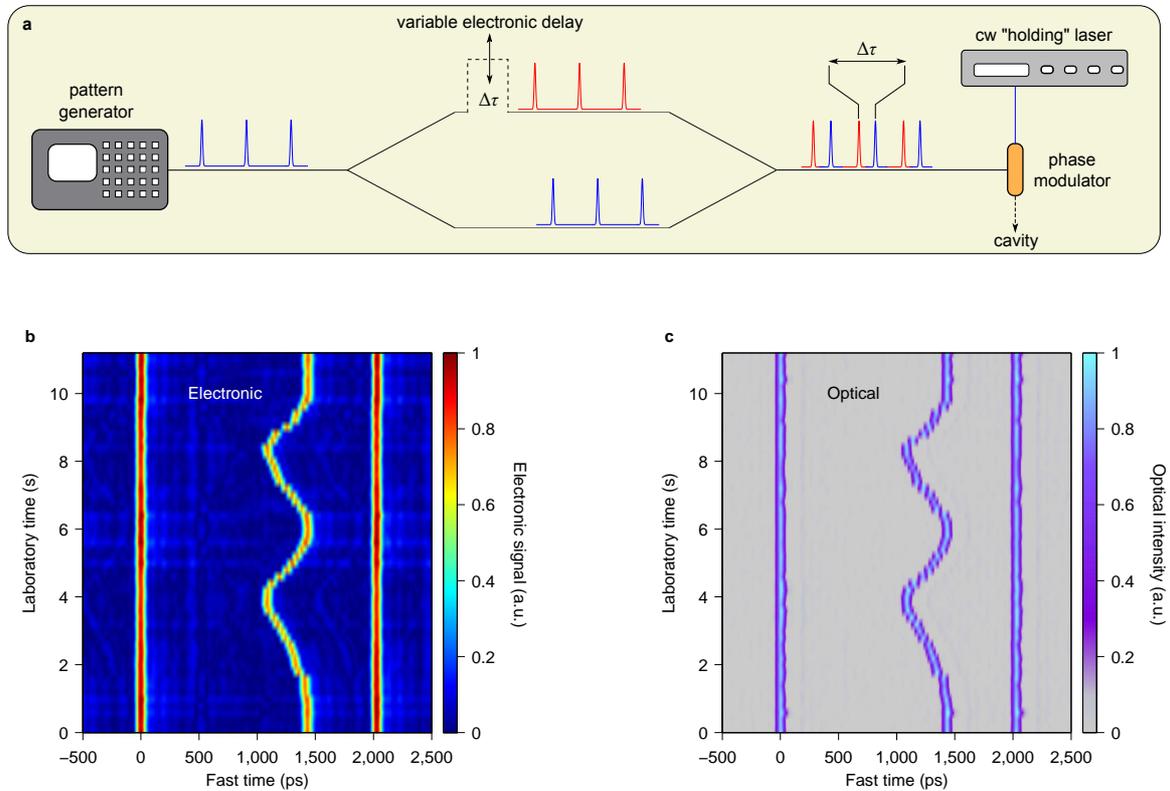}
  \caption{\small \textbf{Demonstration of the temporal tweezing of temporal CSs.} \textbf{a,} Configuration used to
    generate an electronic signal consisting of two interleaved periodic sets of 90~ps quasi-Gaussian pulses whose
    relative delay can be continuously varied. \textbf{b,} Typical real-time manipulation of the electronic signal
    and, \textbf{c,} corresponding identical temporal motion of picosecond optical temporal CSs trapped to the phase
    peaks.}
  \label{manipulation}
\end{figure*}

\subsection{Experimental setup.}

For experimental demonstration, we use a cavity made up of 100~m of standard optical fibre (see Methods). A
1550~nm-wavelength cw laser actively-locked near a cavity resonance generates the holding beam. An external
electro-optic modulator patterns the phase of the holding beam before its launch into the cavity. The electronic (RF)
signal used to drive the modulator is shaped by various components depending on the experiment, as detailed below.

Our configuration supports temporal CSs of $2.6$~ps duration \cite{jang_ultraweak_2013}. We excite them incoherently
through cross-phase modulation between the cw intracavity field and ultrashort pulses picked from the output of a
separate mode-locked ``writing'' laser at a different wavelength \cite{leo_temporal_2010, jang_ultraweak_2013}. This
writing beam is only used once at the start of each measurement: it is coupled into the cavity using a
wavelength-division-multiplexer (WDM), and the WDM also ensures that the writing pulses exit the cavity after a
single roundtrip. Once excited, the CSs persist by themselves in the cavity and all the light of the writing laser is
blocked.

The CS dynamics is monitored by extracting one percent of the intracavity light at each roundtrip for analysis with a
fast photodetector and a real-time digital oscilloscope.

\subsection{Phase modulation trapping.}

We first demonstrate how phase modulation enables robust trapping of temporal CSs. Figure~\ref{PM1}a illustrates the
behaviour of three temporal CSs, initially excited with an 800~ps relative separation, when propagating in the cavity
in the absence of any phase modulation. The colour plot is made up of a vertical concatenation of oscilloscope
recordings of the temporal intensity of the light leaving the cavity at each roundtrip, and reveals, from bottom to
top, how the three CSs evolve over subsequent roundtrips (the temporal resolution of our photodetector is about
50~ps, hence the figure does not capture how short the $2.6$~ps temporal CSs truly are). As can be seen, the temporal
separations between the CSs slowly change and the initial bit pattern distorts over time. This occurs because of
acoustic-mediated interactions as previously reported \cite{jang_ultraweak_2013}.

In contrast, when repeating the experiment in presence of a 10~GHz sinusoidal phase modulation on the holding beam,
Fig.~\ref{PM1}b, no sign of interactions or environmental jitter is observed. The temporal CSs are precisely trapped
to the modulation. Not only does this result confirm the phase modulation trapping scheme, but it shows that acoustic
interactions do not hinder the potential of temporal CSs for optical buffering applications \cite{leo_temporal_2010,
jang_ultraweak_2013}. A concrete example of the latter is illustrated in Fig.~\ref{PM1}c. Here we show the 10~GHz bit
pattern 1001010 1000001 1000101, corresponding to the 7-bit-ASCII representation of ``JAE,'' successfully buffered as
temporal CSs over 2~minutes (i.e. for more than 200 million cavity roundtrips).

\subsection{Demonstration of temporal tweezing.}

Next, we demonstrate the temporal tweezing of trapped CSs. Here, instead of a sinusoidal modulation, we use the
configuration depicted in Fig.~\ref{manipulation}a to generate a re-configurable phase pattern. A pattern generator
produces an electronic signal consisting of quasi-Gaussian pulses with 90~ps duration repeating every 2~ns. That
signal is split into two. One of the split signals experiences an additional variable delay $\Delta\tau$ imparted by
an electronic delay line, before it is recombined with the other. The resulting signal is fed to the phase modulator.
In this way, we imprint onto the holding beam two identical interleaved sets of phase peaks whose relative temporal
separation can be continuously adjusted.

\begin{figure}[t]
  \includegraphics[width = 0.95\columnwidth]{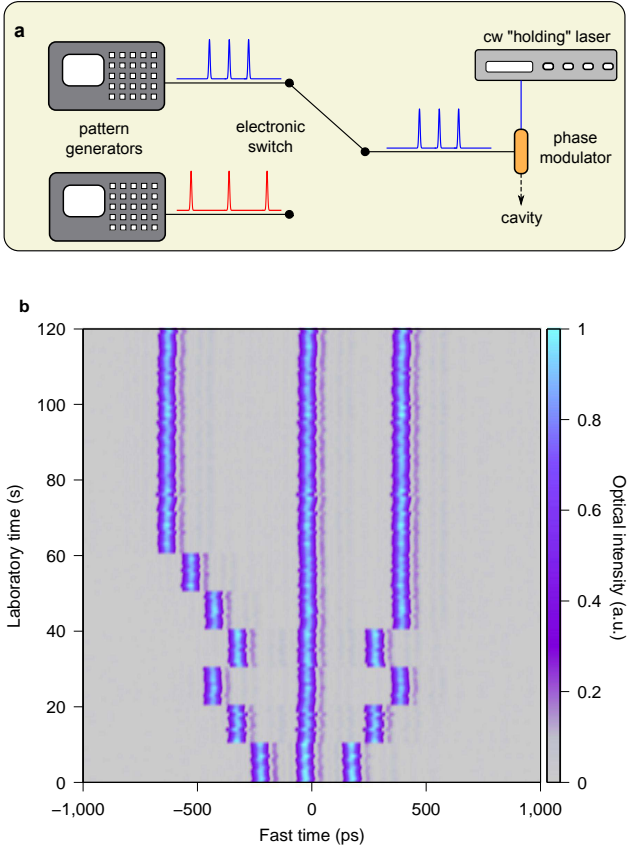}
  \caption{\small \textbf{Discrete tweezing of temporal CSs.} \textbf{a,} By alternative switching and reprogramming
    of two synchronized pattern generators, we achieve, \textbf{b,} discrete simultaneous and independent tweezing of
    multiple temporal CSs.}
  \label{manipulation2}
\end{figure}

Figure~\ref{manipulation}b illustrates the manipulation of the phase pattern. Using a vertical concatenation of
oscilloscope recordings as in Fig.~\ref{PM1}, we show here the \textit{electronic signal} applied to the phase
modulator over time while imposing, by hand, an arbitrary change of the variable delay. The first and the third phase
peaks at 0~ps and 2,000~ps, respectively, correspond to consecutive bits of the pattern generator whilst the middle
one, initially at 1,400~ps, is the delayed replica of the first. As the delay is continuously varied, the middle
phase peak is translated back and forth over a 360~ps range. At the start of this experiment, we also excite three
temporal CSs and ensure they are trapped by adjacent peaks of the initial phase pattern. The evolution of the
temporal \textit{optical intensity} profile of these CSs was recorded simultaneously with the electronic phase signal
shown in Fig.~\ref{manipulation}b, and the result is plotted in Fig.~\ref{manipulation}c. Remarkably, the middle CS
tracks precisely, in real time, the changes of the phase pattern, demonstrating a selective dynamic temporal shift
--- or temporal tweezing --- of an ultrashort picosecond light pulse. We must stress that the magnitude of the shift
we can impart on our temporal CSs is only limited by the cavity roundtrip time. The 360~ps range demonstrated in
Fig.~\ref{manipulation}c corresponds to 140~pulse widths.

To further highlight the flexibility of our scheme, Fig.~\ref{manipulation2} illustrates temporal tweezing in
discrete steps. Here manipulation is performed by alternatively switching the phase modulator electronic feed between
two distinct (but synchronized) 10~GHz pattern generators (Fig.~\ref{manipulation2}a). The patterns are successively
reprogrammed at each step to provide independent and simultaneous temporal tweezing of multiple CSs between the
100~ps bit slots of the generators (Fig.~\ref{manipulation2}b). A completely reconfigurable optical buffer is a
natural outcome of this demonstration. Note that here we have used longer 140~ps electronic pulses, so as to
guarantee the trapping of the temporal CSs even after shifting the corresponding phase pulses by 100~ps.

\begin{figure}[t]
  \includegraphics[width = 0.96\columnwidth]{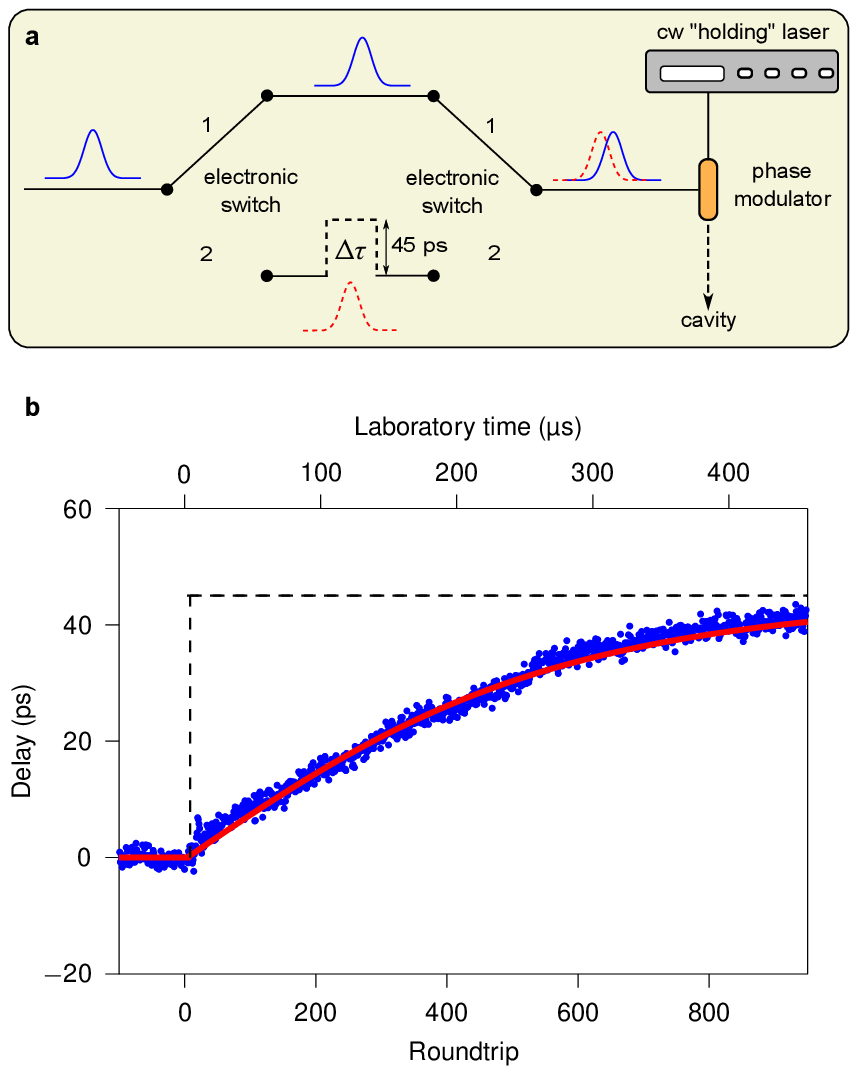}
  \caption{\small \textbf{Transient phase-modulation attraction dynamics.} \textbf{a,} Electronic circuit allowing
    a phase pulse trapping a CS to be abruptly delayed by 45~ps.
    \textbf{b,} Delay experienced by the CS with respect to its initial position (experiment,
    blue dots, and theory, red curve) as it is attracted towards the shifted phase peak over 950~roundtrips (or
    460~$\mu$s). The first 100~roundtrips (with negative numbers) demonstrate the stable trapping of the CS before
    the switches are activated. The dashed line indicates the maximum of the phase pulse.}
  \label{transient}
  \vskip-1.5mm
\end{figure}

\subsection{Transient attraction dynamics.}

The experimental results above clearly demonstrate how phase modulation of the holding beam permits temporal tweezing
of picosecond CSs. These measurements are, however, performed over time-scales that do not allow transient CS
attraction dynamics to be analysed. In particular, we do not resolve in Fig.~\ref{manipulation2}b the time-scale over
which the CSs are attracted to their new position; instead the displacement appears instantaneous.

To address this point, we use the configuration depicted in Fig.~\ref{transient}a. A pattern generator is set to
generate one 110~ps-long phase pulse per cavity roundtrip time. Before feeding these pulses into the phase modulator,
they are sent along one of two different paths set up between two electronic switches, with one of the paths
providing a 45~ps extra delay with respect to the other. The experiment is initiated with a single temporal CS
trapped to the phase peak. The electronic switches are then abruptly and simultaneously activated, which causes the
phase profile to be delayed by 45~ps relative to the CS (switching occurs in about 5~ns, i.e. much faster than the
cavity roundtrip time). As a result, the CS finds itself down the phase pulse, approximately where the slope is the
steepest, and begins to drift towards the shifted phase peak. We monitor the drift by triggering the real-time
oscilloscope from the same signal that controls the electronic switches, and by acquiring a long real-time trace of
the CS train that exits the cavity over the roundtrips immediately following the switching. From the time series we
then infer the drift rate by direct comparison with a simultaneously measured reference set by the electronic signal
driving the phase modulator (see also Methods).

Experimental results are shown as blue dots in Fig.~\ref{transient}b. Here we plot the delay of the CS relative to
its initial temporal position, with the zero roundtrip point marking the activation of the electronic switches. The
first data points (with negative roundtrip numbers) sit stably at 0~ps, and they simply demonstrate the trapping
before the phase pulse is temporally shifted. For positive roundtrip numbers we can see the CS being progressively
delayed, as it is attracted towards the shifted phase peak, initially with a rate of about 75~fs per roundtrip. To
compare with theory, we have directly calculated the expected delay accumulated over successive roundtrips by
iteratively applying Eq.~\eqref{drift} with experimental parameters (see Methods). The prediction is shown as the red
solid line in Fig.~\ref{transient}b. It is in excellent agreement with experimental observations, confirming our
theoretical considerations.

\subsection{Asynchronous phase modulation.}

In the previous experiments, the phase modulation frequency was carefully adjusted to be an integer harmonic of the
cavity free-spectral-range (FSR). This is not, however, a stringent requirement, and temporal tweezing is reliable
even if this condition is not exactly met. In this case, the CS trapping positions are simply slightly offset from
the phase maxima, in direct analogy with the offset of particles from the waist of the trapping beam in conventional
optical tweezers \cite{grier_revolution_2003}. The theory presented in Supplementary Section SII gives the following
condition for the CS trapping position $\tau_\mathrm{CS}$ (measured with respect to a phase maximum):
\begin{equation}
  \phi'(\tau_\mathrm{CS}) = \frac{\Delta f}{\beta_2 L \cdot  \mathrm{FSR} \cdot  f_\mathrm{PM}},
  \label{asyn_pos}
\end{equation}
where $f_\mathrm{PM} = {N\cdot\mathrm{FSR}} + \Delta f$ is the phase modulation frequency and $\Delta f$ its offset
from the closest harmonic of the FSR. Significantly, a limiting value for the frequency mismatch that the trapping
process can tolerate can be derived from the above:
\begin{equation}
  |\Delta f| < |\beta_2| L \cdot  \mathrm{FSR} \cdot  f_\mathrm{PM}\,|\phi'(\tau)|_\mathrm{max} \equiv \Delta f_\mathrm{max}.
  \label{asyn_lim}
\end{equation}
Here $|\phi'(\tau)|_\mathrm{max}$ is the maximum  of the phase gradient, and $\phi(\tau)$ was assumed time-symmetric
for brevity.

\begin{figure}[t]
  \includegraphics[width = 0.95\columnwidth]{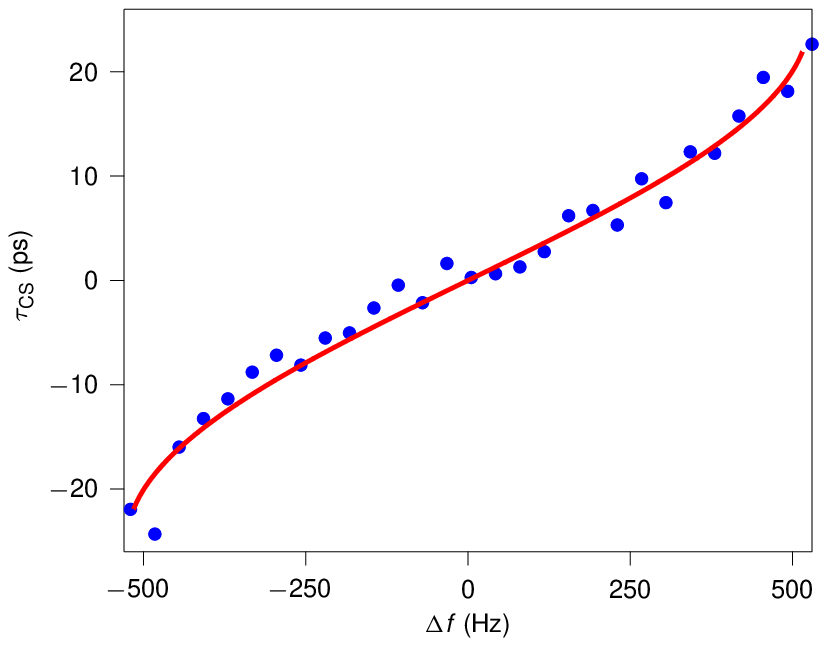}
  \caption{\textbf{Effect of asynchronous phase modulation.} The blue dots show
    experimentally measured temporal position of the stably trapped CS relative to the nearest peak of the sinusoidal
    phase modulation for a range of frequency mismatch $\Delta f$. The red solid curve displays theoretical prediction (see also Methods)
     and agrees well with the experimental data.}
    \vskip-1.5mm
  \label{dtdf}
\end{figure}

We tested this prediction by trapping a single cavity soliton on a sinusoidal phase profile whose frequency was
approximately set to 10~GHz ($\sim 4{,}819~\mathrm{FSR}$). We then systematically adjusted the frequency of the
modulation so as to scan the offset-frequency $\Delta f$. For each value, we compared the measured photodetector
signal to the electronic signal driving the phase-modulator, which allowed us to deduce the position of the CS
relative to the nearest phase peak (see Methods). Experimental results are shown as the blue dots in Fig.~\ref{dtdf},
and are in excellent agreement with the theoretical prediction (red solid line) derived from Eq.~\eqref{asyn_pos}. We
find an experimental trapping range of $\Delta f = \pm 534$~Hz. Beyond this limit, the CSs are no longer trapped;
instead we observe that their relative separation against the RF signal varies from roundtrip-to-roundtrip. This
value again matches very well the theoretical limit $\Delta f_\mathrm{max} = 530~\mathrm{Hz}$ calculated from
Eq.~(\ref{asyn_lim}) for our experimental conditions, again validating our analysis.

\subsection{Discussion.}

The scheme demonstrated in this work enables distortionless trapping and manipulation of picosecond temporal cavity
solitons. On the one hand, the trapping mechanism allows for long-range soliton interactions
\cite{jang_ultraweak_2013} and environmental jitter to be suppressed, which directly permits all-optical buffering
for extended periods of time. On the other hand, the ability to dynamically control the pulse positions by
manipulating the phase profile of the cavity holding beam renders the buffer fully reconfigurable. All the
experimental observations agree completely with theory and numerical simulations (Supplementary Section SI).

Our experiment is not particularly sensitive to any parameters. Of particular significance is the fact that the phase
modulation does not have to be accurately synchronized with the cavity FSR, with the observed tolerable frequency
mismatch of $\pm 0.5$~kHz being sufficiently large to avoid active locking of the phase modulator to the cavity. The
speed at which the CS temporal positions can be manipulated is limited by the drift rate [Eq.~\eqref{drift}]. In our
experiments, shifting over a single bit slot in a 10~GHz sequence occurs over approximately $500\ \mu\mathrm{s}$.
Faster manipulation could be obtained by using fibres with larger group-velocity dispersion or phase modulation with
larger maximum gradient; no attempts have been made at optimization. Miniaturization of the technology could also be
possible by harnessing the ability of monolithic microresonators to support temporal CSs~\cite{herr_temporal_2014}.

On a more general level, our work illustrates how the unique particle-like characteristics of solitons
\cite{zabusky_interaction_1965, hasegawa_transmission_1973} can be leveraged to achieve all-optical control of light.
Specifically, we have demonstrated arbitrary selective manipulation of the temporal positions of ultrashort optical
pulses within a train, with no limitation on the range of time delays over which pulses can be translated. In doing
so, we have effectively realized a temporal tweezer for light, in a platform fully compatible with existing
fibre-based communication technologies. These results could have significant implications for all-optical information
processing.

\section*{Methods}

\begingroup

\small

\subsection{Basic experimental setup.}

For our experiments, we use a continuous-wave (cw) pumped passive fibre cavity set-up similar to that used in
Ref.~\citenum{jang_ultraweak_2013}. The cavity is made up of $L = 100$~m of standard single-mode silica optical fibre
(Corning SMF-28) with measured group-velocity dispersion $\beta_2 = -21.4~\mathrm{ps^2\,km^{-1}}$ and nonlinearity
coefficient $\gamma = 1.2~\mathrm{W^{-1}\,km^{-1}}$ at a wavelength of 1550~nm. The fibre is laid in a ring
configuration with a 90/10 input coupler closing the loop. Stimulated Brillouin scattering
\cite{agrawal_nonlinear_2006} is prevented with the inclusion in the ring of a fibre isolator with 60~dB extinction.
Overall, the cavity has a free-spectral range (FSR) of $2.075$~MHz (or roundtrip time $t_\mathrm{R} =
\mathrm{FSR}^{-1} \simeq 0.48\ \mu\mathrm{s}$) and a finesse $\mathcal{F} = 21.5$. The cavity is coherently driven
with a 1550~nm-wavelength Koheras AdjustiK{\tiny\texttrademark{}} E15 distributed-feedback fibre laser (linewidth $<
1$~kHz). The 20~mW cw laser output can be amplified up to 1~W before being coupled into the cavity through the input
coupler to act as the ``holding'' beam. The optical frequency of the laser is actively locked at a set detuning from
a cavity resonance with a commercial 100~kHz proportional-integral-derivative (PID) controller (SRS SIM960). The
error signal of the PID controller is simply obtained by comparing the power reflected off the cavity with a
reference level.

The writing pulses used to excite the temporal CSs via cross-phase-modulation are emitted by a 1532~nm-wavelength
picosecond mode-locked laser with a 10~GHz repetition rate. Two intensity modulators, one driven by a 10~GHz
electronic pattern generator, the other by a gating pulse, are used in sequence to select a single realization of a
desired pattern of writing pulses.

To trap and manipulate the temporal CSs, the holding beam is phase-modulated using a telecommunications electro-optic
phase-modulator with a 10~GHz bandwidth. The phase-modulator is placed before the cavity input coupler, between the
AdjustiK laser and the amplifier. The electronic (RF) signal used to drive the modulator is shaped by various RF
components depending on the experiment and originates from either a sinusoidal signal generator or an electronic
pattern generator. The pulses from our pattern generator are 90~ps long (full-width at half-maximum, FWHM), and
longer pulses can be obtained when needed with an extra RF filter. In all experiments, the CS dynamics and our
temporal manipulations are monitored by extracting one percent of the intracavity power through an additional fibre
tap coupler incorporated into the ring cavity. The extracted light is detected and analysed with a $12.5$~GHz
amplified PIN photodiode connected to a 40~GSa/s real-time oscilloscope. This system has a $\sim 50$~ps impulse
response which sets the temporal resolution of the fast-time measurements.

\subsection{Oscilloscope acquisition rate.}

The colour plots made up of a vertical concatenation of oscilloscope recordings of the temporal intensity profile of
the cavity output have been obtained with a 1~frame/s acquisition rate for Figs.~\ref{PM1}a--c and
Fig.~\ref{manipulation2}b, while a rate of 5~frame/s has been used for Figs.~\ref{manipulation}b--c.

\subsection{Transient attraction experiment.}

The experimental results shown in Fig.~\ref{transient} are measured by triggering the real-time oscilloscope from
the same TTL signal that controls the electronic switches, and by acquiring a long real-time signal of the CS train
that exits the cavity over roundtrips immediately following the switching. Since the oscilloscope has a limited
memory depth of 2~million points at the highest sampling rate of 40~GSa/s, data can only be acquired about
100~roundtrips at a time. This is not sufficient to reveal the full dynamics. To circumvent this issue, we acquired
10~independent sets of data where the oscilloscope's acquisition window with respect to the TTL trigger is
systematically delayed. These sets are then carefully combined so as to obtain a full recording of CS drift dynamics
over about 1,000~roundtrips, as shown in Fig.~\ref{transient}.

To obtain the theoretical comparison, we used experimental cavity parameters and inferred the functional form of the
phase profile $\phi(\tau)$ by measuring the electronic pulse driving the modulator. It was found to have a smooth
Gaussian shape with 110~ps FWHM and an amplitude of $2.7$~rad. The theoretical prediction was then obtained by
iteratively applying Eq.~\eqref{drift} over consecutive roundtrips.

\subsection{Asynchronous phase-modulation experiment.}

In experiments examining the effect of asynchronous phase modulation, the frequency of the sinusoidal electronic
signal driving the phase modulator was systematically adjusted in steps of 37.5~Hz. For each value, we simultaneously
measured the photodetector signal at the cavity output as well as the electronic signal driving the modulator. The
temporal position of the CS relative to the nearest phase maximum was then obtained by comparing the two recorded
signals. Since these signals invariably possess different electrical path lengths, they come with a fixed
instrumental temporal delay, preventing direct measurement of the absolute separation. To overcome this, we assume
that for perfect synchronisation ($\Delta f = 0$) the separation is zero, as predicted by theory, and we justify this
assumption \emph{a posteriori} by the excellent agreement with theoretical predictions seen for all $\Delta f$. To
obtain the $\Delta f = 0$ reference, we exploit the fact that, for a symmetric phase profile, $\Delta f = 0$ occurs
midway between the trapping limit frequencies $\pm \Delta f_\mathrm{max}$. They are easy to find since for $|\Delta
f|>\Delta f_\mathrm{max}$ the CS is not trapped and its separation relative to the RF signal changes from
roundtrip-to-roundtrip.

Since this experiment uses (co)sinusoidal phase modulation, the theoretical prediction for the CS position
$\tau_\mathrm{CS}$ can be expressed analytically. Specifically, substituting $\phi(\tau) = A\cos(2\pi
f_\mathrm{PM}\tau)$ into Eq.~\eqref{asyn_pos} we derive:
\begin{equation}
  \tau_\mathrm{CS} = -\frac{1}{2\pi f_\mathrm{PM}}
                      \sin^{-1}\left[\frac{\Delta f}{2\pi A\beta_2 L\cdot\mathrm{FSR}\cdot f_\mathrm{PM}^2}\right].
  \label{asyn_pos2}
\end{equation}
We note that the experimental data shown in Fig.~\ref{dtdf} exhibit the expected inverse-sine characteristics. The
theoretical prediction shown as the red solid line in this figure has been derived from the above expression, using
the experimental phase modulation frequency $f_\mathrm{PM}\simeq 10$~GHz and amplitude $A = 0.19$~rad. For the
tolerable frequency mismatch, we have $|\phi'(\tau)|_\mathrm{max}= 2\pi f_\mathrm{PM} A$, yielding $\Delta
f_\mathrm{max} = 530$~Hz.

\bigskip

\section*{Acknowledgments}

\noindent This work was supported by the Marsden Fund Council (government funding), administered by the Royal Society
of New Zealand. J.K.J. also acknowledges the support of a University of Auckland Doctoral Scholarship.

\section*{Author Contributions}

\noindent J.K.J. performed all the experiments and numerical simulations. M.E. clarified the theory of cavity soliton
trapping and manipulation. S.G.M. supervised the experimental work and, together with J.K.J., designed the
experiments. M.E. and S.C. wrote the manuscript. S.C. and S.G.M. supervised the overall project. All authors
contributed to discussing and interpreting the results.

\section*{Additional information}

\noindent Supplementary information accompanies this paper. Correspondence and requests for materials should be
addressed to J.K.J. and S.G.M.

\section*{Competing financial interests}

\noindent The authors declare no competing financial interests.

\endgroup

\makeatletter\close@column@grid\makeatother

\clearpage

\relpenalty=9999 \binoppenalty=9999

% Supplementary figure and equation numbers start with "S"
\renewcommand{\theequation}{S\arabic{equation}}
\renewcommand{\thefigure}{S\arabic{figure}}

\setcounter{equation}{0}\setcounter{figure}{0}

\makeatletter

% Section title ragged right
\def\section{%
  \@startsection
    {section}%
    {1}%
    {\z@}%
    {0.6cm \@plus1ex \@minus .2ex}%
    {0.3cm}%
    {%
      \normalfont\small\bfseries
      \raggedright
    }%
}%

\makeatother

\title{Supplementary Information\\ Temporal tweezing of light: trapping and manipulation of temporal cavity solitons}

\author{Jae K. Jang}
\author{Miro Erkintalo}
\author{St\'ephane Coen}
\author{Stuart G. Murdoch}

\affiliation{Department of Physics, The University of Auckland, Private Bag 92019, Auckland 1142, New Zealand}

\begin{abstract}
  \noindent This article contains supplementary theoretical information to the manuscript entitled ``Temporal
  tweezing of light: trapping and manipulation of temporal cavity solitons.'' Specifically, we derive the rate of
  phase-modulation induced drift using the mean-field Lugiato-Lefever equation, and show how tweezing can be
  reproduced in full numerical simulations. We also derive the equations describing the impact of asynchronous
  phase modulation.
\end{abstract}

\thispagestyle{preprint}

\maketitle

\section{Theory of temporal tweezing}

\noindent Temporal tweezing relies on picosecond CSs exhibiting an attractive time-domain drift towards the maxima of
the intracavity phase profile. In the main manuscript, we explain the physics underlying the attraction in terms of
the CSs shifting their instantaneous frequencies in reaction to a phase modulation. Here we elaborate on this aspect
by presenting a more formal mathematical treatment.

The dynamics of light in a high-finesse passive fibre cavity is governed by the mean-field Lugiato-Lefever equation
(LLE) \cite{lugiato_spatial_1987_supp, haelterman_dissipative_1992_supp},
\begin{multline}
  t_\mathrm{R}\frac{\partial E(t,\tau)}{\partial t} =
    \bigg[ -\alpha - i \delta_0 -iL\frac{\beta_2}{2}\frac{\partial^2}{\partial \tau^2} \\
  + i\gamma L |E|^2 \bigg] E + \sqrt{\theta}\,E_\mathrm{in}.
  \label{LL}
\end{multline}
Here $t$ is the slow time describing the evolution of the intracavity field envelope $E(t,\tau)$ over subsequent
cavity roundtrips, while $\tau$ is a fast time describing the temporal profile of the field envelope in a reference
frame travelling at the group velocity of the holding beam in the cavity. $E_\mathrm{in}$ is the field of the holding
beam with power $P_\mathrm{in}=|E_\mathrm{in}|^2$. The parameter $\alpha = \pi/\mathcal{F}$ accounts for all the
cavity losses, with $\mathcal{F}$ the cavity finesse. Denoting $\phi_0$ as the linear phase-shift acquired by the
intracavity field over one roundtrip with respect to the holding beam, $\delta_0 = 2\pi l - \phi_0$ measures the
phase detuning of the intracavity field to the closest cavity resonance (with order $l$). Finally, $L$ is the cavity
length, $\beta_2$ and $\gamma$ are, respectively, the dispersion and nonlinear coefficient of the fibre, and $\theta$
is the input coupler power transmission coefficient.

To  analyse the effect of a phase-modulation of the holding field, we follow the approach of
Ref.~\citenum{firth_optical_1996-1_supp}. Specifically, assuming a phase-modulation temporal profile $\phi(\tau)$ that
repeats periodically with a period equal to the cavity roundtrip time~$t_\mathrm{R}$ (or an integer fraction of it),
we can write $E_\mathrm{in}(\tau)=F_\mathrm{in}\exp[i\phi(\tau)]$, where $F_\mathrm{in}$ is a constant scalar.
Substituting this expression into Eq.~\eqref{LL}, together with the ansatz ${E(t,\tau)=F(t,\tau)\exp[i\phi(\tau)]}$,
yields
\begin{multline}
  t_\mathrm{R}\frac{\partial F(t,\tau)}{\partial t} -\beta_2 L \phi' \frac{\partial F}{\partial \tau}
    = \bigg[ -\alpha_\mathrm{F} - i \delta_\mathrm{F} -iL\frac{\beta_2}{2}\frac{\partial^2}{\partial \tau^2}\\
  + i\gamma L |F|^2 \bigg] F + \sqrt{\theta}\,F_\mathrm{in}\,.
  \label{LL2}
\end{multline}
Here $\alpha_\mathrm{F} = \alpha-\beta_2L\phi''/2$ and $\delta_\mathrm{F} = \delta_0-\beta_2L(\phi')^2/2$, with
$\phi' = \mathrm{d}\phi/\mathrm{d}\tau$ and $\phi'' = \mathrm{d^2}\phi/\mathrm{d}\tau^2$. We first note that this
equation, together with the ansatz above, makes clear that the cw intracavity field on which the CSs are superimposed
has the same phase modulation as that imposed on the external holding beam (to within a constant phase shift). Next,
we consider the second term on the left-hand-side of Eq.~\eqref{LL2} which gives rise to a local change of the
group-velocity of the field. Indeed, with the change of variable ${\tau \rightarrow \tau_\mathrm{F} = \tau +
\beta_2L\phi' t/t_\mathrm{R}}$ (and assuming that the phase modulation is slow in comparison to the duration of CSs),
the second term on the left-hand-side cancels out and Eq.~\eqref{LL2} can be recast into a form identical to the
LLE~\eqref{LL}, but with a time-independent holding field $F_\mathrm{in}$. Such an equation, in the $\tau_\mathrm{F}$
frame, admits stationary CS solutions \cite{firth_optical_1996-1_supp}. The CS solutions of Eq.~\eqref{LL2} thus drift in
the fast time domain~$\tau$, and move with respect to the phase modulation, with a rate
\begin{equation}
  V_\mathrm{drift} = \frac{\mathrm{d}\tau}{\mathrm{d}t} = -\frac{\beta_2 L \phi'}{t_\mathrm{R}}.
\label{Vdrift}
\end{equation}
$V_\mathrm{drift}$ is generally small enough for the total change of the temporal position of a CS over one roundtrip
to be approximated as $\tau_\mathrm{drift} \sim V_\mathrm{drift}\,t_\mathrm{R} = -\beta_2L\phi'$. In our experiments
$\beta_2<0$ (anomalous dispersion), $\tau_\mathrm{drift} = |\beta_2|L\phi'$, and a CS overlapping with an increasing
part of the phase modulation profile ($\phi'>0$) will be temporally delayed ($\tau_\mathrm{drift}>0$) whilst another
overlapping with a decreasing part ($\phi'<0$) will be advanced ($\tau_\mathrm{drift}<0$). Both scenarios result in
CSs approaching the phase maxima, where the drift ceases ($\phi' = 0 \Rightarrow \tau_\mathrm{drift} = 0$). It is
worth noting that $\tau_\mathrm{drift}$ coincides exactly with the temporal delay accrued by a signal with an
instantaneous angular frequency shift $\delta\omega = -\phi'$, propagating through a length $L$ of fibre whose
group-velocity dispersion is $\beta_2$ \cite{agrawal_nonlinear_2006_supp}. Therefore the above LLE-analysis is fully
consistent with the simple physical description presented in our main manuscript.

\begin{figure*}[t]
  \includegraphics[width = \textwidth, clip = true]{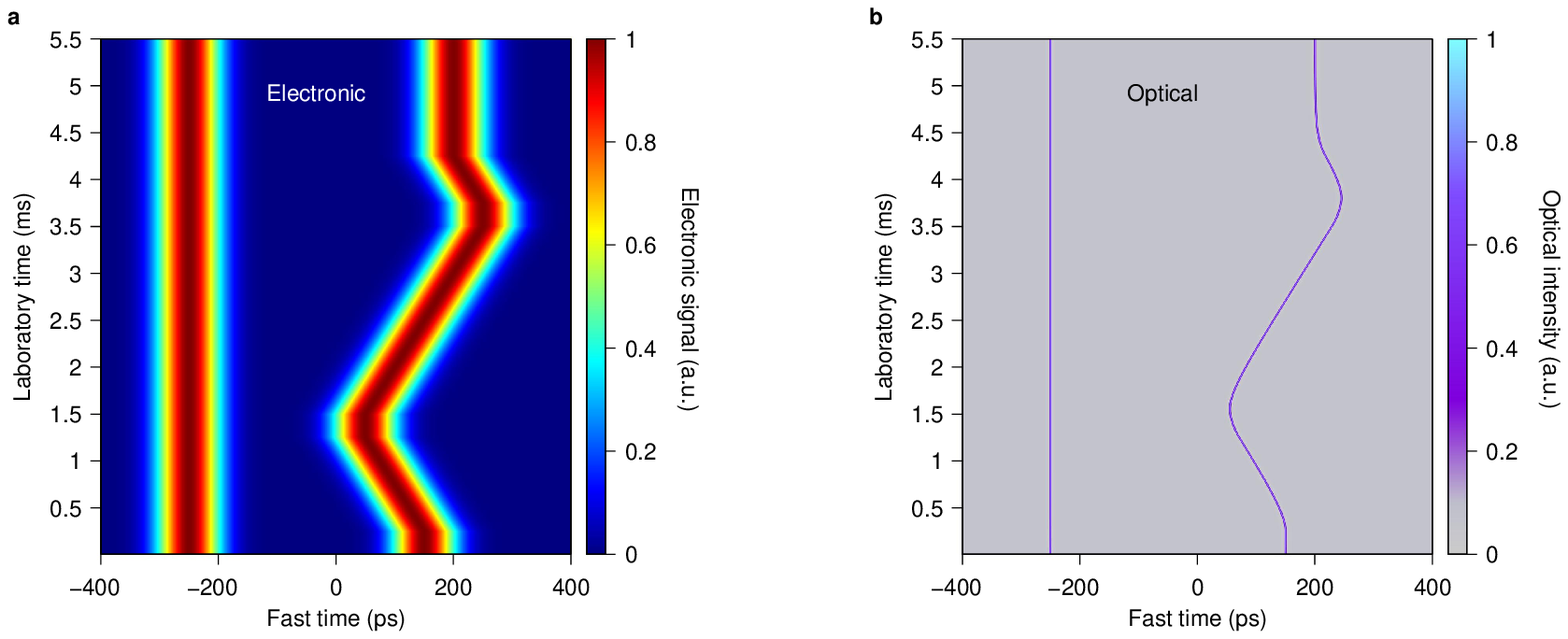}
  \caption{\small \textbf{Illustrative numerical simulations of temporal tweezing.} \textbf{a,} Phase profile
    $\phi(\tau,t)$ used in the simulations to illustrate the temporal tweezing of CSs. The second phase peak shifts
    back and forth in sequence at a rate of $\pm\,50$~fs/roundtrip with intermediate stationary periods.
    \textbf{b,} The corresponding dynamical evolution of two $2.6$~ps-long temporal CSs. The trailing CS
    always drifts towards the shifting phase maximum. These simulations use experimental parameters, with
    $t_\mathrm{R} = 0.48~\mu s$, $\alpha = 0.146$, $\delta_0 = 0.41$~rad, $L = 100$~m, $\beta_2 = -21.4\ \mathrm{ps^2\,km^{-1}}$,
    $\gamma = 1.2\ \mathrm{W^{-1}\,km^{-1}}$, $\theta = 0.1$, $P_\mathrm{in} = 960~\mathrm{mW}$. The phase peaks are
    Gaussian with 90~ps FWHM and $2.0$~rad amplitude.}
  \label{sim1}
  \vskip -3mm
\end{figure*}
If the phase modulation $\phi(\tau)$ is dynamically varied (such that $\phi(\tau,t)$ also depends on the slow-time
variable $t$) then the intracavity phase profile will adjust accordingly, with a response time governed by the cavity
photon lifetime~$t_\mathrm{ph}$ (with our parameters $t_\mathrm{ph}\sim 3.4~t_\mathrm{R}$). This is the principle of
our temporal tweezer that allows trapped CSs to be temporally shifted, as is demonstrated experimentally in the main
manuscript. Temporal tweezing of CSs can also be readily studied using direct numerical simulations of
Eq.~\eqref{LL}. In Fig.~\ref{sim1} we show an example of numerical results where we start with two CSs at the top of
two 90~ps wide (FWHM) Gaussian-shaped phase pulses initially separated by 400~ps. After 500 roundtrips, we start
shifting the second phase pulse at a rate of 50~fs per roundtrip, first bringing it 100~ps closer, then moving it
away and back again until we eventually stop. We can see how the corresponding CS follows these manipulations. At the
end of the sequence, it gets trapped again stably at the phase peak. Note that all the parameters used in this
simulation (listed in the caption of Fig.~\ref{sim1}) match values used in our experiments. The only difference is
the much faster manipulation of the phase profile, performed over a much smaller number of roundtrips, to expedite
the computations and to facilitate visualization of the transient dynamics. In particular, we remark that the
simulation here captures the actual two orders of magnitude difference between the durations of the CSs ($\sim 2.6
~\mathrm{ps}$) and the RF pulses ($\sim 90~\mathrm{ps}$), whilst the experiments reported in our main manuscript are
subject to limited bandwidth of the detection electronics.

\section{Asynchronous phase-modulation}

\noindent Here we show theoretically that CS trapping can be achieved even if the phase modulation frequency is not
an exact multiple of the cavity FSR. In particular, we derive the equation that allows the CS trapping position to be
predicted.

Consider phase modulation with a frequency $f_\mathrm{PM} = {N\cdot\mathrm{FSR}} + \Delta f$ slightly mismatched with
respect to the $N$th harmonic of the FSR. If $\Delta f>0$, the intracavity field takes effectively too long to
complete one cavity roundtrip and be in synchronism with the phase modulation. The temporal delay that the
intracavity field accumulates with respect to the phase profile of the holding beam over one roundtrip is given by
$\Delta\tau = t_\mathrm{R} - N T_\mathrm{PM}$, with $T_\mathrm{PM} = 1/f_\mathrm{PM}$ the period of the phase
modulation. Given that $t_\mathrm{R} = \mathrm{FSR}^{-1}$, and assuming $|\Delta f|\ll \mathrm{FSR}$, we have
$\Delta\tau \simeq \Delta f/[\mathrm{FSR} \cdot f_\mathrm{PM}]$. A CS can remain effectively trapped provided that,
during each roundtrip, the phase-modulation induced drift exactly cancels the relative delay accrued from the
synchronization mismatch: $\tau_\mathrm{drift} + \Delta\tau = 0$. Recalling that $\tau_\mathrm{drift} = -\beta_2
L\phi'$, we can see that in the presence of a non-zero frequency mismatch $\Delta f$, the CSs will not be trapped to
the peak of the phase profile (where the gradient $\phi'$ vanishes). Rather, they will be trapped at a temporal
position $\tau_\mathrm{CS}$ along the phase profile that satisfies the equation:
\begin{equation}
  \phi'(\tau_\mathrm{CS}) = \frac{\Delta f}{\beta_2 L \cdot  \mathrm{FSR} \cdot  f_\mathrm{PM}}.
  \label{asyn_pos_supp}
\end{equation}
This also allows us to obtain the limiting value for the frequency mismatch that the trapping process can tolerate:
\begin{equation}
  |\Delta f| < |\beta_2| L \cdot  \mathrm{FSR} \cdot  f_\mathrm{PM}\,|\phi'(\tau)|_\mathrm{max} \equiv \Delta f_\mathrm{max},
\label{asyn_lim_supp}
\end{equation}
where $|\phi'(\tau)|_\mathrm{max}$ is the maximum  of the gradient of the phase profile, and for brevity we have
assumed $\phi(\tau)$ to be symmetric. When $|\Delta f| = \Delta f_\mathrm{max}$, the CSs will be trapped to the
steepest point along the phase profile, but when $|\Delta f| > \Delta f_\mathrm{max}$ the effective temporal drift
arising from the frequency mismatch is too large to be overcome by the phase-modulation attraction.

\section{Asynchronous phase-modulation in the LLE model}

\noindent Asynchronous phase-modulation can also be analysed in the mean-field LLE by considering a driving field
modulated with a travelling phase profile. Transferring the intracavity field into a reference frame where the phase
pattern is stationary, $\xi = \tau+(\Delta\tau/t_\mathrm{R}) t$, and injecting
$E_\mathrm{in}=F_\mathrm{in}\exp[i\phi(\xi)]$ and $E=F\exp[i\phi(\xi)]$ into Eq.~\eqref{LL} we obtain:
\begin{align}
  \label{LL3}
  &t_\mathrm{R}\frac{\partial F(t,\tau)}{\partial t} -(-\Delta\tau+\beta_2 L \phi') \frac{\partial F}{\partial \xi}\\
    =& \bigg[ -\alpha_\mathrm{F} - i \delta_\mathrm{FF} -iL\frac{\beta_2}{2}\frac{\partial^2}{\partial \xi^2}\nonumber
   + i\gamma L |F|^2 \bigg] F + \sqrt{\theta}\,F_\mathrm{in},
\end{align}
where $\delta_\mathrm{FF} = \delta_0-\beta_2L(\phi')^2/2+\Delta\tau\phi'$ and $\phi' = \mathrm{d}\phi/\mathrm{d}\xi$.
It can again be seen that the CS solutions will be stationary in the reference-frame moving with the intracavity
phase profile, provided that the second term on the left-hand side vanishes. This occurs when the solitons are
located at a temporal coordinate $\xi$ that solves the equation:
\begin{equation}
  \phi'(\xi) = \frac{\Delta\tau}{\beta_2 L} = \frac{\Delta f}{\beta_2 L \cdot  \mathrm{FSR} \cdot  f_\mathrm{PM}}.
\end{equation}
This result coincides with that  in Eq.~\eqref{asyn_pos_supp}, derived using simple physical arguments.

\end{document}